\documentclass[showpacs,preprintnumbers,amsmath,amssymb]{revtex4}
\usepackage{amsmath,amssymb,graphics,epsfig,subfigure}
\usepackage{color}

\begin{document}
\renewcommand{\baselinestretch}{1.3}

\title{Merger estimates for rotating Kerr black holes in modified gravity}

\author{Shao-Wen Wei \footnote{weishw@lzu.edu.cn},
        Yu-Xiao Liu \footnote{liuyx@lzu.edu.cn}}

\affiliation{Research Center of Gravitation, Lanzhou University, Lanzhou 730000, China\\
             Institute of Theoretical Physics, Lanzhou University, Lanzhou 730000, China\\
             Key Laboratory for Magnetism and Magnetic of the Ministry of Education, Lanzhou University, Lanzhou 730000, China}

\begin{abstract}
In this paper, we explore the signatures of non-rotating and rotating black hole mergers in the matter-free modified gravity. First, we solve the unstable circular null orbits and the innermost stable circular timelike orbits via the geodesic motion. The characteristic quantities of these orbits are systematically analyzed by varying the black hole spin and the scalar field parameter of the gravity. Then based on it, we study the ringdown modes from the light ring/quasinormal modes correspondence. The final spins of the merged black holes are also estimated with the Buonanno-Kidder-Lehner recipe. Several black hole merging cases are investigated in detail. All these results show that the black hole mergers are closely dependent of the scalar field parameter of the gravity.
\end{abstract}



\maketitle

\section{Introduction}

Recently, gravitational waves have been directly observed by LIGO and Virgo \cite{Abbott}, and have opened a new window to probe the fundamental questions of our Universe and gravitational theory. As expected, these gravitational waves were produced by merging binary black holes or neutron stars. Particularly, the exact determination of the masses, spins, and other parameters for the initial and final black holes will greatly improve our test near the strong gravitational regimes. Moreover, different modified gravities are also expected to be tested or restricted by using the gravitational wave observations \cite{KonoplyaZh,Bambij}.

Although recent observations incline to support general relativity (GR) \cite{AbbottBP,Yunes}, some subtle potential deviations may be witness in the coming years when more merging events are detected and higher improved signal-to-noise is achieved. In order to discriminate different relevant gravity theories by the observations, one needs first to examine their differences. Importantly, it is necessary to identify the key signatures of the waveforms of the black hole mergers for these theories, which is also very useful to exactly determine the dynamic properties and the nature of these initial and final black holes.

On the other hand, there is an interesting modified gravity (MOG), the scalar-tensor-vector gravity theory proposed by Moffat \cite{Moffat0}. Besides the metric tensor fields, extra massive vector field and scalar fields were introduced. The scalar fields strengthen the gravitational attraction, while the vector field produces an effective repulsive gravitational force. Such theory can be treated as an alternative to GR for dealing with the galaxy rotation curves and galaxy clusters without introducing the dark matter \cite{Rahvar,MoffatR,Toth}.

Particularly, many recent works were devoted to explore the novel properties of the nonrotating and rotating MOG black holes \cite{Moffat3}. Black hole shadow was briefly examined in Ref. \cite{Moffat4}. The thermodynamics of these black holes were studied in Ref. \cite{Mureika}. The geodesics and accretion disk were investigated in Refs. \cite{Lee,Sharif,Hussain,Perez,Sheoran}. Misaligned spin merging black holes were discussed in Ref. \cite{Moffat5}. And the quasinormal modes (QNMs) were calculated in Ref. \cite{Manfredi}. All these results imply that there exists a significant difference between MOG and GR.

In the present work, we would like to consider the Kerr-MOG black hole merger, and two issues of the merger are studied. The first one is the gravitational wave emission patterns at the last stage, i.e., the ringdown stage, of the merger. In the eikonal limit, the QNMs are associated with the unstable circular null geodesics, the light ring, of the black hole in the asymptotically flat spacetime (for recent progress, see Refs. \cite{Cardoso,Glampedakis,Kokkotas}). Although there exists a small deviation between the numerical data \cite{Khanna,Konoplya,Salcedo,Khoo}, it is still a good approximation.  According to this light ring/QNM correspondence, the real part of the QNM is related to the angular velocity of the massless particle orbiting around the light ring. And its imaginary part is related to the Lyapunov exponent of the light ring. So through studying the light ring from the null geodesics, we can obtain the ringdown modes of the black hole merger. This has a theoretical guidance to distinguish this MOG from GR by probing the merger dynamics at the ringdown stage.

Another issue of the merger we will pursue is the estimate of the final black hole spin. For this section, we adopt the Buonanno-Kidder-Lehner (BKL) recipe \cite{Buonanno}, which is based on the angular momentum and mass conservations. And such recipe has gained a great success on estimating the final black hole spin. Moreover, this approach has two advantages. First, it considers the orbital angular momenta and both initial spins of the merged black holes. Second, it can be applied to general mass ratio mergers. Then through solving the innermost stable circular orbits (ISCO) from the timelike geodesics, one can obtain the final spin of the black hole following this BKL recipe. The study has been carried out for the Kerr black hole and Einstein-Maxwell-dilaton black hole mergers. Comparing with the numerical simulations, the result shows that the BKL formula is quite accurate \cite{Buonanno,Chatrabhuti}. More importantly, such approach can also be easily generalized to other different merger configurations, and some interesting and novel dynamics might be explored immediately. For example, the ``flip'' phenomenon was discussed with such recipe. Moreover, it was found that it is impossible to spin-up a black hole to extremal values through merger scenarios \cite{Buonanno}. However, considering the energy loss, a near extreme spinning black hole can be produced by equal mass maximally spinning aligned mergers \cite{Kesden}.

This paper is organized as follows. In Sec. \ref{KerrMOG}, we give a brief review of the black hole metric and its geodesics. Based on the geodesics, we calculate the radius and minimum impact parameter for the light rings. Further, we study the QNMs at the ringdown stage in Sec. \ref{ringdown}. In Sec. \ref{merger}, we first numerically calculate the ISCO. Then based on the result, we follow the BKL recipe and estimate the final black hole spins for several merger configurations. Finally, the conclusions are presented in Sec. \ref{Conclusion}.

\section{Kerr-MOG black hole and geodesics}
\label{KerrMOG}

The action of the scalar-tensor-vector gravity is formulated as \cite{Moffat0}
\begin{eqnarray}
 S=S_{GR}+S_{\phi}+S_{S}+S_{M},
\end{eqnarray}
with
\begin{eqnarray}
 S_{GR}&=&\frac{1}{16\pi}\int d^{4}x\sqrt{-g}\frac{R}{G},\\
 S_{\phi}&=&\int d^{4}x\sqrt{-g}\left(-\frac{1}{4}B^{\mu\nu}B_{\mu\nu}
     +\frac{1}{2}\mu^{2}\phi^{\mu}\phi_{\mu}\right),\\
 S_{S}&=&\int d^{4}x\sqrt{-g}\frac{1}{G^{3}}
  \left(\frac{1}{2}g^{\mu\nu}\nabla_{\mu}G\nabla_{\nu}G-V(G)\right)
  +\int d^{4}x\frac{1}{\mu^{2}G}
  \left(\frac{1}{2}g^{\mu\nu}\nabla_{\mu}\mu\nabla_{\nu}\mu-V(\mu)\right),
\end{eqnarray}
where $\phi^{\mu}$ is a Proca type massive vector field with mass $\mu$, $G(x)$ and $\mu(x)$ are two scalar fields, and their potentials are respectively $V(G)$ and $V(\mu)$. $S_{M}$ denotes the matter action. The tensor field $B_{\mu\nu}=\partial_{\mu}\phi_{\nu}-\partial_{\nu}\phi_{\mu}$, and it satisfies the following equations
\begin{eqnarray}
 &&\nabla_{\nu}B^{\mu\nu}=0,\\
 &&\nabla_{\sigma}B_{\mu\nu}+\nabla_{\mu}B_{\nu\sigma}+\nabla_{\nu}B_{\sigma\mu}=0.
\end{eqnarray}
The energy momentum tensor for the vector field is
\begin{eqnarray}
 T_{\phi\mu\nu}&=&-\frac{1}{4\pi}\left(B_{\mu}^{\;\sigma}B_{\nu\sigma}
      -\frac{1}{4}g_{\mu\nu}B^{\sigma\beta}B_{\sigma\beta}\right).
\end{eqnarray}
Since the effect of the mass $\mu$ of the vector field displays at kiloparsec scales from the source, it can be neglected for a black hole solution. Moreover, we can also treat $G$ as a constant independent of the spacetime coordinates. With this hypothesis, the action will be simplified and easy to deal with,
\begin{eqnarray}
 S=\int d^{4}x\sqrt{-g}\left(\frac{R}{16\pi G}-\frac{1}{4}B^{\mu\nu}B_{\mu\nu}\right).
\end{eqnarray}
The corresponding field equation reads
\begin{eqnarray}
 G_{\mu\nu}=-8\pi G T_{\phi\mu\nu},\label{field}
\end{eqnarray}
and $G$ is corresponded to the Newton's gravitational constant as $G=G_{N}(1+\alpha)$ with $\alpha$ being a dimensionless parameter. So $\alpha$ can be used to measure the deviation of MOG from GR.

The rotating Kerr-MOG black hole solution can be obtained by solving the field equation (\ref{field}). In Boyer-Lindquist coordinates, the metric is given by \cite{Moffat3}
\begin{eqnarray}
 ds^{2}=-\frac{\Delta}{\rho^{2}}\big(dt-a\sin^{2}\theta d\phi\big)^{2}
        +\frac{\rho^{2}}{\Delta}dr^{2}+\rho^{2}d\theta^{2}
        +\frac{\sin^{2}\theta}{\rho^{2}}\Big(adt-(r^{2}+a^{2})d\phi\Big)^{2},
\end{eqnarray}
with the metric functions given by
\begin{eqnarray}
 \Delta&=&r^{2}-2G_{N}(1+\alpha)mr+a^{2}+m^{2}G_{N}^{2}\alpha(1+\alpha),\\
 \rho^{2}&=&r^{2}+a^{2}\cos^{2}\theta.
\end{eqnarray}
For simplicity, we adopt $G_{N}=1$. The parameter $m$ is related to the ADM mass $M$ of the black hole as \cite{Sheoran}
\begin{eqnarray}
 M=(1+\alpha)m.
\end{eqnarray}
Solving $\Delta=0$, we can obtain the radius of the black hole horizon
\begin{eqnarray}
 r_{\pm}=M\pm \sqrt{\frac{M^{2}}{1+\alpha}-a^{2}}.
\end{eqnarray}
It is clear that such black hole can possess two horizons for $M^{2}>(1+\alpha)a^{2}$, one degenerate horizon for $M^{2}=(1+\alpha)a^{2}$, and no horizon related to naked singularity for $M^{2}<(1+\alpha)a^{2}$. So the black hole has a maximum spin
\begin{eqnarray}
 \frac{a_{max}}{M}=\frac{1}{\sqrt{1+\alpha}}.\label{extremal}
\end{eqnarray}
It is worth noting that this Kerr-MOG black hole will reduce to the Kerr black hole when $\alpha=0$. On the other hand, if we set $a=0$, it will be a static black hole with two horizons $r_{\pm}=M(1\pm 1/\sqrt{1+\alpha})$. Further taking $\alpha=0$, the Schwarzschild black hole solution will be recovered.

In this rotating Kerr-MOG background, the geodesics of a test particle with unit mass is \cite{Moffat3}
\begin{eqnarray}
 \dot{t}&=&\frac{g_{\phi\phi}E+g_{\phi t}L}{g_{\phi t}^{2}-g_{tt}g_{\phi\phi}},\\
 \dot{\phi}&=&-\frac{g_{tt}L+g_{t\phi}E}{g_{t\phi}^{2}-g_{tt}g_{\phi\phi}}.\\
 \rho^{2}\dot{r}&=&\sigma_{r}\sqrt{\mathcal{R}},\label{rmotion}\\
 \rho^{2}\dot{\theta}&=&\sigma_{\theta}\sqrt{\Theta},
\end{eqnarray}
where
\begin{eqnarray}
 \mathcal{R}&=&\Big(aL-E(r^{2}+a^{2})\Big)^{2}
         -\bigg(a^{2}+r^{2}-2Mr+\frac{\alpha M^{2}}{1+\alpha}\bigg)
               (\mathcal{K}+\mu^{2}r^{2}),\\
 \Theta&=&\mathcal{K}-a^{2}\mu^{2}\cos^{2}\theta-(aE\sin^{2}\theta-L)^{2}\csc^{2}\theta.
\end{eqnarray}
The sign functions $\sigma_{r}=\pm$ and $\sigma_{\theta}=\pm$ are independent from each other. And $\mathcal{K}$ is the carter constant. The values of $\mu^{2}$ are 1 and 0 for massive particle and photon, respectively. Here, we consider the motion limited in the equatorial plane. So we have $\theta=\frac{\pi}{2}$ and $\mathcal{K}=(aE-L)^{2}$. Then we can reexpress Eq. (\ref{rmotion}) as the following form
\begin{eqnarray}
 \dot{r}^{2}+V_{eff}=0,
\end{eqnarray}
where the effective potential is given by
\begin{eqnarray}
 V_{eff}=-\frac{\mathcal{R}}{\rho^{4}}.
\end{eqnarray}

\section{Ringdown modes}
\label{ringdown}

The final stage of a black hole merger is known as the ringdown stage. In this stage, the final black hole settles to a stationary one, and it is characterized by linearized vibrational modes, i.e., the QNMs. In the eikonal limit, such mode is associated with the spacetime's light ring. And this light ring/QNM correspondence is effective for an asymptotically flat spacetime. The modes of $n=|m|$ are considered to be the most powerful emitters during the gravitational waves and thus are most easily detectable by observatories, so we mainly focus on them in this paper. Fortunately, these modes are also the easiest ones to fit the eikonal approximation. More interestingly, these modes with positive or negative $m$ are related with the equatorial prograde or retrograde orbital motion.

\begin{figure}
\center{\subfigure[]{\label{Rpsa}
\includegraphics[width=6cm]{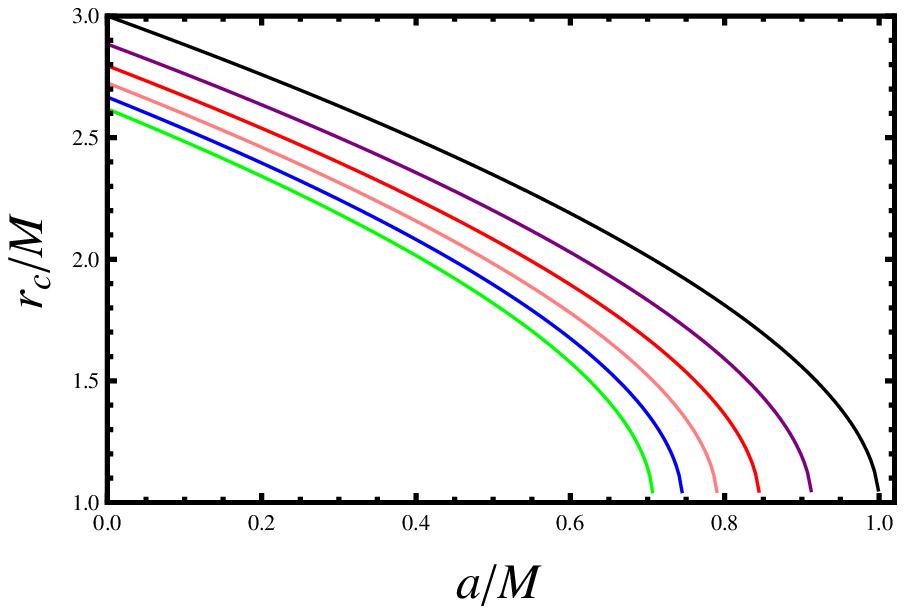}}
\subfigure[]{\label{Upsa}
\includegraphics[width=6cm]{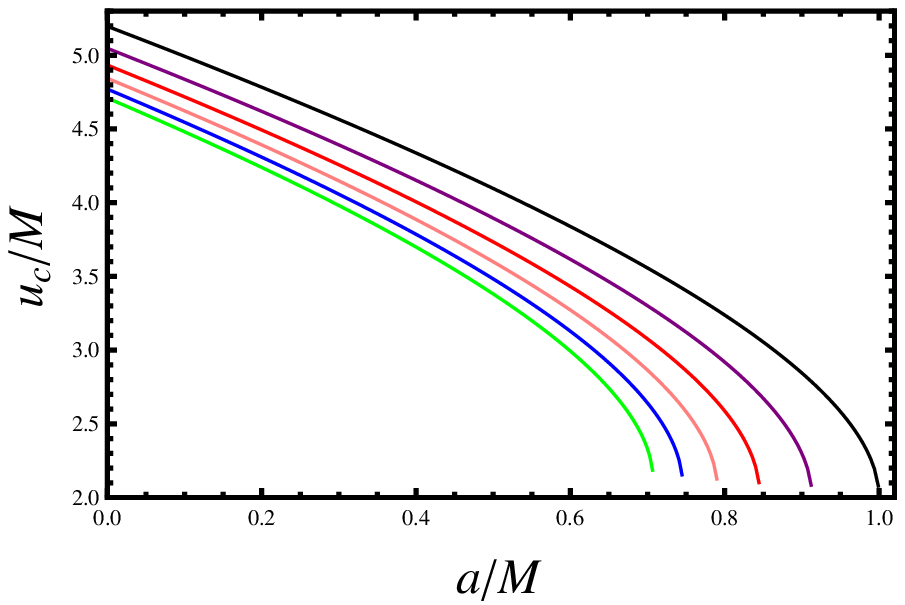}}}
\caption{Radius and minimum impact parameter for the prograde light rings. (a) $r_{c}$ vs. $a$. (b) $u_{c}$ vs. $a$. The parameter $\alpha$=0, 0.2, 0.4, 0.6, 0.8, 1 from top to bottom.}\label{pUpsa}
\end{figure}

\begin{figure}
\center{\subfigure[]{\label{Rpsb}
\includegraphics[width=6cm]{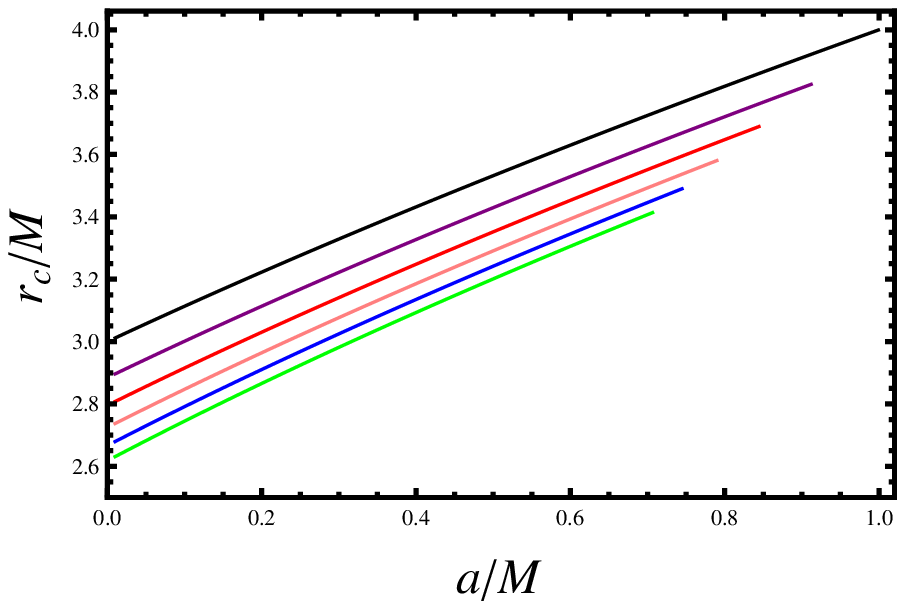}}
\subfigure[]{\label{Upsb}
\includegraphics[width=6cm]{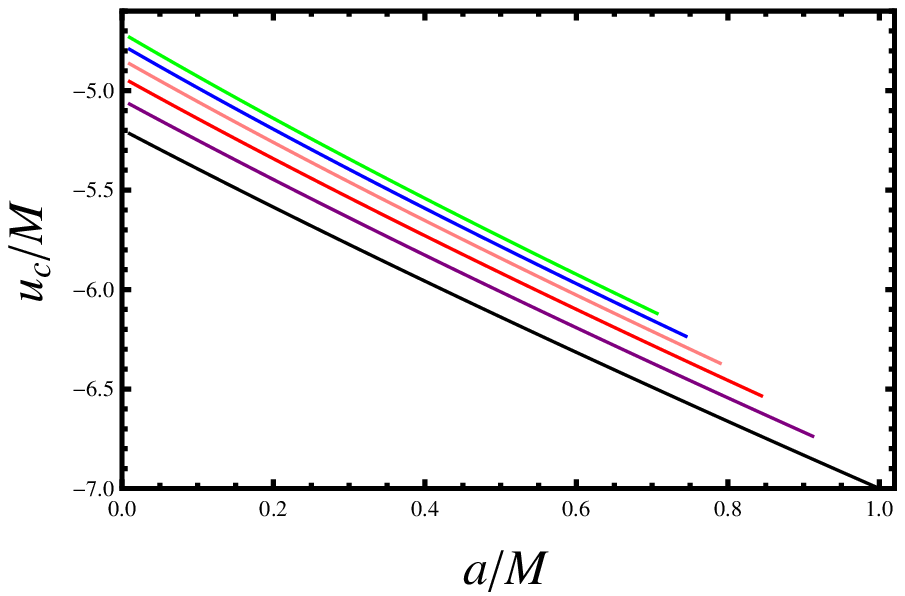}}}
\caption{Radius and minimum impact parameter for the retrograde light rings. (a) $r_{c}$ vs. $a$ with $\alpha$=0, 0.2, 0.4, 0.6, 0.8, 1 from top to bottom. (b) $u_{c}$ vs. $a$ with $\alpha$=0, 0.2, 0.4, 0.6, 0.8, 1 from bottom to top.}\label{pUpsb}
\end{figure}

These modes are expressed in the following form \cite{Cardoso,Mashhoona,Schutz,IyerWill}
\begin{eqnarray}
 \omega_{\text{QNM}}=\Omega_{c}m-i(n+1/2)|\lambda|,\label{QNM}
\end{eqnarray}
where $m$, $n$, $\Omega_{c}$, and $\lambda$ are the quantum overtone number, angular momentum, angular velocity, and Lyapunov exponent of the light ring, respectively. Moreover $\lambda$ and $\Omega_{c}$ can be calculated with the geodesics
\begin{eqnarray}
 \lambda=\sqrt{-\frac{V_{\text{eff}}''}{2\dot{t}^{2}}}\bigg|_{r_{c}},\quad
 \Omega_{c}=\frac{\dot{\phi}}{\dot{t}}\bigg|_{r_{c}},\label{ee}
\end{eqnarray}
with $r_{c}$ being the radius of the light ring. Here the effective potential for the photon reads
\begin{eqnarray}
 V_{eff}=\frac{(aE-L)^{2}\Delta-(aL-E(a^{2}+r^{2}))^{2}}{\rho^{4}}.
\end{eqnarray}
The characteristic parameters of the light rings are determined by the conditions
\begin{eqnarray}
 V_{eff}=0,\quad V'_{eff}=0,\quad V''_{eff}<0.\label{veffcond}
\end{eqnarray}
Substituting the effective potential into these conditions, one will get
\begin{eqnarray}
 (aL-E(a^{2}+r^{2}))^{2}-(aE-L)^{2}\Delta=0,\\
 4Er(aL-E(a^{2}+r^{2}))-(aE-L)^{2}\Delta'=0,\\
 8E^{2}r^{2}-4E(aL-E(a^{2}+r^{2}))-(aE-L)^{2}\Delta''>0.
\end{eqnarray}
Solving them, we can obtain the radius $r_{c}$ and the minimum impact parameter $u_{c}$ of the light rings. For the nonrotating black hole with $a=0$, we have
\begin{eqnarray}
 \frac{u_{c}}{M}&=& \frac{L}{ME}=\frac{\sqrt{27+45\alpha+17\alpha^{2}-\alpha^{3}+((1+\alpha)(9+\alpha))^{3/2}}}
        {\sqrt{2}(1+\alpha)},\\
 \frac{r_{c}}{M}&=&\frac{3+3\alpha+\sqrt{(1+\alpha)(9+\alpha)}}{2(1+\alpha)}.
\end{eqnarray}
When $\alpha=0$, the above result reduces to the Schwarzschild black hole case, i.e.,
\begin{eqnarray}
 \frac{u_{c}}{M}=3\sqrt{3},\quad \frac{r_{c}}{M}=3.
\end{eqnarray}
On the other hand, for the rotating black hole with $a\neq0$, due to the dragging effect, the light rings will be different for prograde and retrograde cases. Although it is impossible to obtain the analytical formula, we can numerically solve these conditions (\ref{veffcond}). And the results are presented in Figs. \ref{pUpsa} and \ref{pUpsb} for the prograde and retrograde orbits. For the prograde orbit, we find that both $r_{c}$ and $u_{c}$ decrease with the spin $a$ and the parameter $\alpha$. For the retrograde orbit, $r_{c}$ increases with the spin $a$ and decreases with $\alpha$, while $u_{c}$ decreases with $a$ and increases with $\alpha$.

\begin{figure}
\center{\subfigure[]{\label{Omegaa}
\includegraphics[width=6cm]{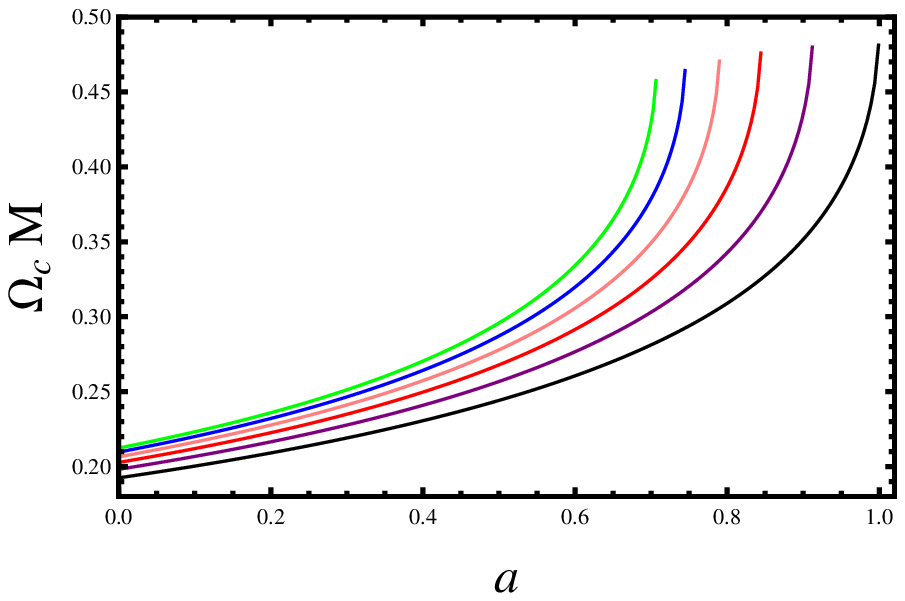}}
\subfigure[]{\label{Lambdaa}
\includegraphics[width=6cm]{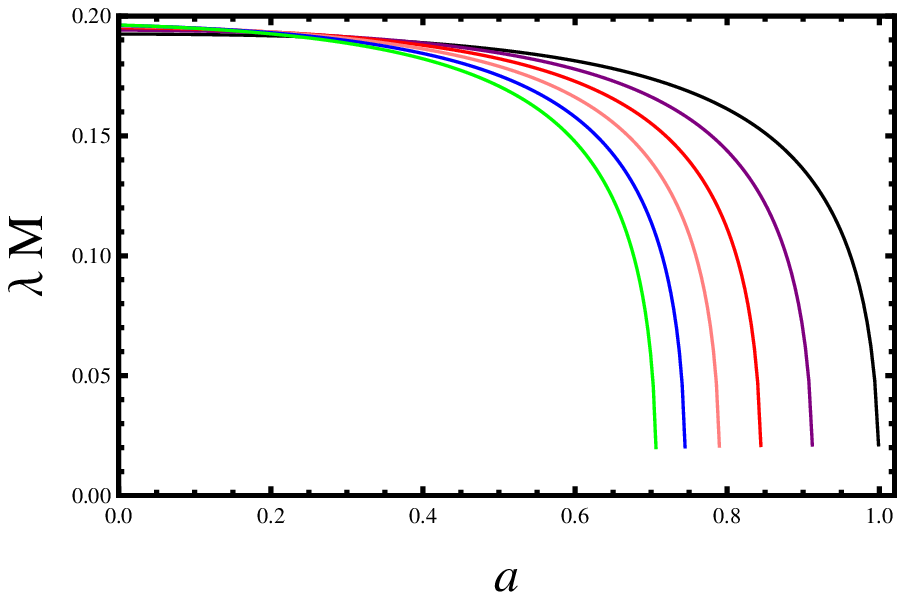}}}
\caption{Angular velocity and Lyapunov exponent for the prograde light rings. (a) $\Omega_{c}$ vs. $a$. (b) $\lambda$ vs. $a$. The parameter $\alpha$=0, 0.2, 0.4, 0.6, 0.8, 1 from right to left.}\label{pLambdaa}
\end{figure}

\begin{figure}
\center{\subfigure[]{\label{Omegab}
\includegraphics[width=6cm]{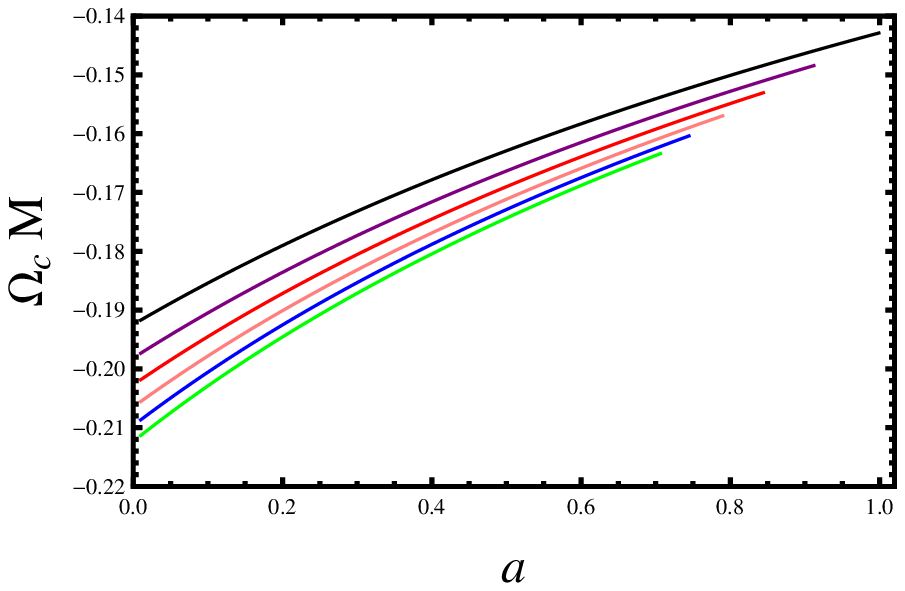}}
\subfigure[]{\label{Lambdab}
\includegraphics[width=6cm]{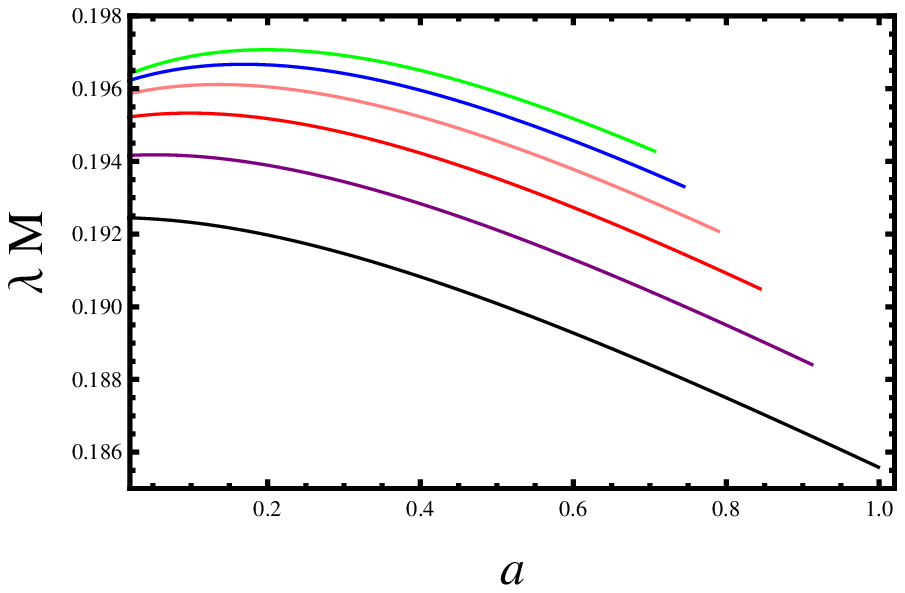}}}
\caption{Angular velocity and Lyapunov exponent for the retrograde light rings. (a) $\Omega_{c}$ vs. $a$ with parameter $\alpha$=0, 0.2, 0.4, 0.6, 0.8, 1 from top to bottom. (b) $\lambda$ vs. $a$ with parameter $\alpha$=0, 0.2, 0.4, 0.6, 0.8, 1 from bottom to top.}\label{pLambdab}
\end{figure}

For the nonrotating black hole, the angular velocity and Lyapunov exponent of the unstable null geodesics are analytically obtained
\begin{eqnarray}
 \Omega_{c}M&=&\frac{\sqrt{2}(1+\alpha)}{\sqrt{27+45\alpha+17\alpha^{2}-\alpha^{3}+((1+\alpha)(9+\alpha))^{3/2}}},\label{pqw1}\\
 \lambda M&=&\frac{2\sqrt{(1+\alpha)^{3}(9+\alpha+\sqrt{9+10\alpha+\alpha^{2}})}}
  {(3+3\alpha+\sqrt{9+10\alpha+\alpha^{2}})^{2}}.\label{pqw2}
\end{eqnarray}
For the rotating black hole, the results are shown in Figs. \ref{pLambdaa} and \ref{pLambdab}. For the prograde orbit, with the increase of spin $a$, the angular velocity $\Omega_{c}$ increases, while the Lyapunov exponent $\lambda$ decreases. Moreover, $\Omega_{c}$ increases with $\alpha$. The Lyapunov exponent $\lambda$ increases with $\alpha$ for small $a$, and decreases with $\alpha$ for large $a$. For the retrograde orbit, $\Omega_{c}$ increases with spin $a$ and decreases with $\alpha$. On the other hand, the Lyapunov exponent $\lambda$ increases with $\alpha$. And for small fixed $\alpha$, $\lambda$ slightly decreases with the spin $a$. However, for large fixed $\alpha$, $\lambda$ first increases, and then decreases with $a$.

Before ending this section, we would like give some notes for the light ring/QNM correspondence. First, our calculation mainly bases on the geodesics, which measures the motion of a pointlike test particle in the background. So it may be in good approximation if one of the black hole is small and other one is huge during the merger. Interestingly, the results given in Refs. \cite{Berti,Sperhake,Tiec} imply that the point particle approximation is also accurate for the equal mass black hole collisions. Thus such correspondence is reasonable on calculating the ringdown modes during the black hole merger.

One the other hand, it is worth to examine the influence of the additional field, i.e., the scalar field parameter $\alpha$, on the light ring/QNM correspondence. As showed in Refs. \cite{Salcedo,Khoo}, the QNMs for the black hole in the Einstein-dilaton-Gauss-Bonnet gravity have a close relation with the coupling parameter. It is also find that the geodesics correspondence predicts only the axial modes of the perturbation. And they are well consistent with each other for small coupling parameter. Here let turn to consider the influence of the parameter $\alpha$ on the light ring/QNM correspondence. For simplicity, we only consider the nonrotating case. Here we list the numerical data in Table \ref{tab1} for the QNMs calculated from the asymptotic iteration method \cite{Manfredi} and from our equations (\ref{pqw1}) and (\ref{pqw2}). From the table, we can see that the real part and the imaginary part of the QNMs have different behaviors with the parameter $\alpha$. For $\alpha=0$, the real part of the QNMs obtained from the two methods highly coincides with each other, and the relative deviation $\delta=0.09\%$. Moreover, the $\delta$ of the real part will increase with $\alpha$. For example, $\delta$ approaches to about $10\%$ for $\alpha=9$. In contrast, the relative deviation of the imaginary part of the QNMs has a slight decrease with $\alpha$, and it is always under $5\%$ for $\alpha=0\sim9$. Combining with the result of the real part and the imaginary part of the QNM, we can arrive the conclusion that the light ring/QNM correspondence is effective for the small parameter $\alpha$. Hence, we only limit our attention for small parameter, i.e., $\alpha\leq1$.

\begin{table}[h]
\begin{center}
\begin{tabular}{ccccc}
  \hline\hline
    & $\alpha=0$ & $\alpha=1$ & $\alpha=4$ & $\alpha=9$ \\\hline
AIM \cite{Manfredi} &0.5779-0.7063i&0.6891-0.7200i&0.7687-0.6988i&0.7927-0.6743i\\\hline
light ring/QNM &0.5774-0.6736i&0.6370-0.6868i&0.6922-0.6753i&0.7177-0.6582i\\\hline
$\delta$(\%) &0.09-4.63i&7.56-4.61i&9.95-3.36i&9.46-2.39i
\\\hline\hline
\end{tabular}
\caption{QNMs for the nonrotating black hole in modified gravity. The first row is the QNMs for the electromagnetic perturbation with $n=|m|=3$ obtained with asymptotic iteration method given in Ref. \cite{Manfredi}. The second row is obtained with Eqs. (\ref{pqw1}) and (\ref{pqw2}). The parameter $\delta$ measures the relative deviation between these two methods.}\label{tab1}
\end{center}
\end{table}

\section{Final spin estimation}
\label{merger}

\subsection{BKL recipe}

Here we would like to briefly review the BKL recipe, which mainly depends on the conservations and is very easily to be understood. First, it considers that the mass of this merged system is conserved. Therefore, the mass of the final black hole equals to the total mass of the initial black holes, i.e.,
\begin{eqnarray}
 M=M_{1}+M_{2}.
\end{eqnarray}
To the first order, this is a good approximation. For example, the total radiated energy remains very small, about $M_{radiated}\sim 5\%M_{initial}$, in the gravitational wave observations \cite{Abbott}.

On the other hand, for a reasonable merger, the two black holes orbit around each other during the first stage, i.e., the inspiral stage. Due to the energy loss via a gravitational wave emission, the binary orbit contracts gradually and the system evolves quasiadiabatically. During this stage, one can assume that the individual spins of the black holes will remain constant. Once the ISCO radius is reached, the orbit becomes unstable. This will lead to a merger, and then a final black hole is formed. At this stage, the radiation of the angular momentum with respect to that of the system is small. So the angular momentum of the system can be treated as a conserved quantity. Moreover, it is justified to estimate the contribution of the orbital angular momentum to the final black hole spin by adopting the orbital angular momentum of a test particle orbiting at the ISCO of the final black hole. Finally, the dimensionless final black hole spin $a_{f}$ can be expressed as \cite{Buonanno}
\begin{eqnarray}
 a_{f}=\frac{\tilde{L}_{orb}(r_{ISCO}, a_{f})}{M}
  +\frac{M_{1}a_{1}}{M}+\frac{M_{2}a_{2}}{M},
\end{eqnarray}
where $\tilde{L}_{orb}(r_{ISCO}, a_{f})$ denotes the orbital angular momentum of a test particle with reduced mass $M_{1}M_{2}/M$ orbiting at the ISCO of the final black hole with spin $a_{f}$. Without loss of generality, we can assume $M_{1}\geq M_{2}$, and then the final spin can also be reexpressed in the following convenient form \cite{Buonanno}
\begin{eqnarray}
 a_{f}=L(r_{ISCO}, a_{f})\nu
  +\frac{\chi_{1}M}{4}(1+\sqrt{1-4\nu})^{2}
  +\frac{\chi_{2}M}{4}(1-\sqrt{1-4\nu})^{2},\label{BKL}
\end{eqnarray}
where $\chi_{i}=a_{i}/M_{i}$ is the dimensionless spins of the initial black holes and $L(r_{ISCO}, a_{f})$ is the angular momentum of a unit mass test particle. The mass parameter $\nu=\frac{M_{1}M_{2}}{M^{2}}$ and has a value in the range of $[0, 1/4]$. By solving the timelike geodesics, we can obtain the radius and the orbital angular momentum of the ISCO. Then with given the initial mass parameter $\nu$ and spins $\chi_{1}$ and $\chi_{2}$, the final black hole spin can be calculated with Eq. (\ref{BKL}). Importantly, since the parameter $\alpha$ deforms the gravitational constant charactered by the spacetime background, we fix $\alpha$ before and after the black hole merger.

\subsection{ISCO}

\begin{figure}
\center{\subfigure[]{\label{Lpro}
\includegraphics[width=6cm]{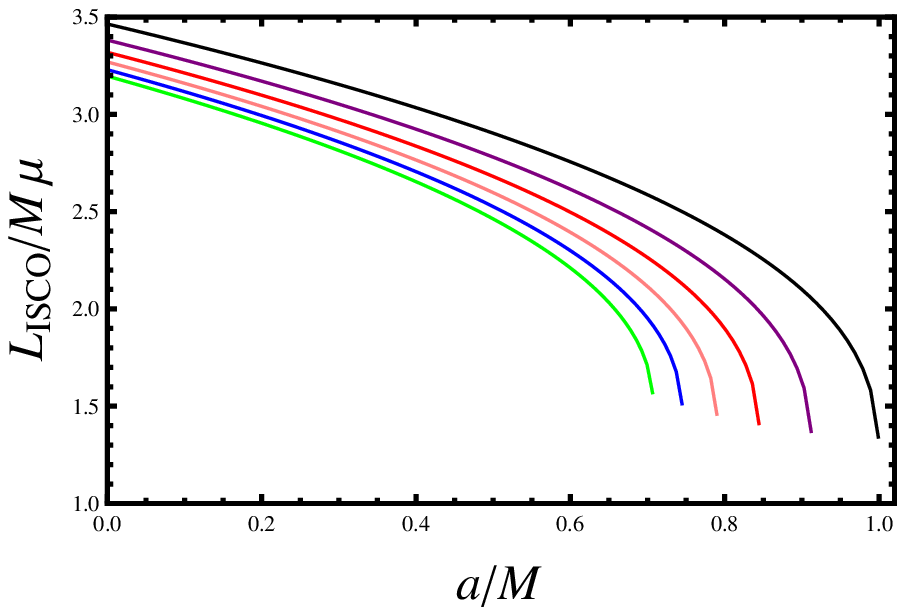}}
\subfigure[]{\label{Rpro}
\includegraphics[width=6cm]{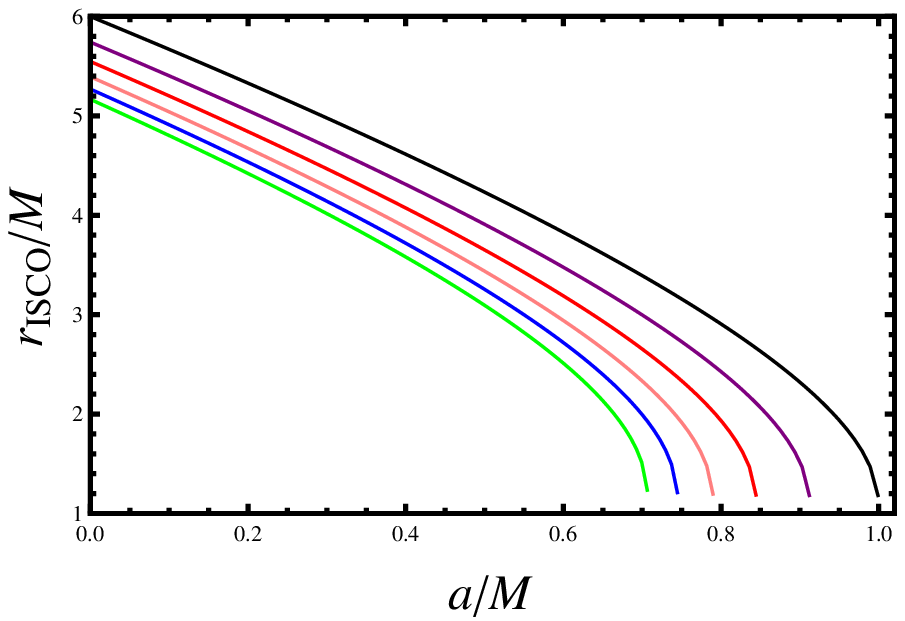}}}
\caption{Angular momentum and radius for the prograde ISCO. (a) $L_{ISCO}$ vs. $a$. (b) $r_{ISCO}$ vs. a. Parameter $\alpha$=0, 0.2, 0.4, 0.6, 0.8, 1 from top to bottom.}\label{pRpro}
\end{figure}

\begin{figure}
\center{\subfigure[]{\label{Lret}
\includegraphics[width=6cm]{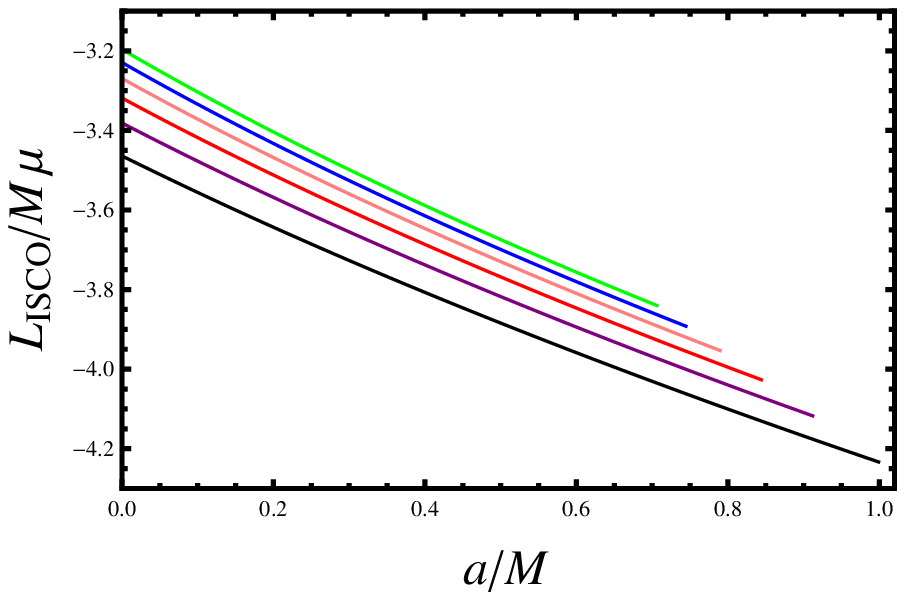}}
\subfigure[]{\label{Rret}
\includegraphics[width=6cm]{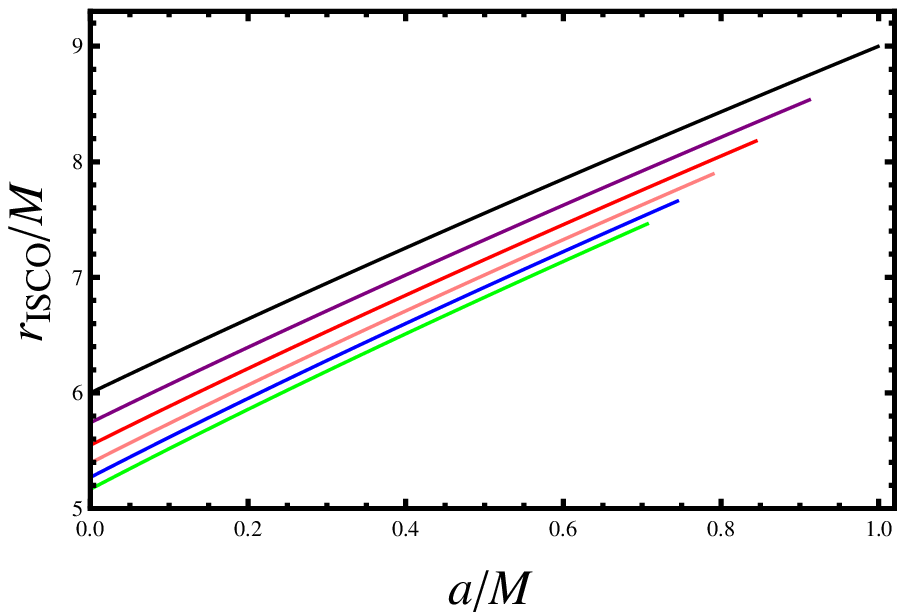}}}
\caption{Angular momentum and radius for the retrograde ISCO.  (a) $L_{ISCO}$ vs. $a$ with $\alpha$=0, 0.2, 0.4, 0.6, 0.8, 1 from bottom to top. (b) $r_{ISCO}$ vs. $a$ with $\alpha$=0, 0.2, 0.4, 0.6, 0.8, 1 from top to bottom.}\label{pRret}
\end{figure}

As mentioned above, in order to obtain the final spin of the black hole, we first need to solve the ISCO. So here we will consider such special orbit for a test particle. From the geodesics, the conditions to determine the ISCO are
\begin{eqnarray}
 V_{eff}=0,\quad \partial_{r}V_{eff}=0, \quad \partial_{r,r}V_{eff}=0.\label{ISCOcondition}
\end{eqnarray}
Plunging the effective potential, the conditions reduce to
\begin{eqnarray}
 (aL-E(a^{2}+r^{2}))^{2}-((aE-L)^{2}+\mu^{2} r^{2})\Delta=0,\\
 4Er(aL-E(a^{2}+r^{2}))+2\mu^{2}r\Delta+((aE-L)^{2}+\mu^{2} r^{2})\Delta'=0,\\
 4E(aL-E(a^{2}+r^{2}))-8E^{2}r^{2}+2\mu^{2}\Delta+4\mu^{2}r\Delta'+((aE-L)^{2}+\mu^{2} r^{2})\Delta''=0.
\end{eqnarray}
For the nonrotating black hole, we can express the angular momentum and energy in terms of the radius of the ISCO as
\begin{eqnarray}
 L/\mu&=&\pm\frac{r\sqrt{M(r-M\alpha+r\alpha)}}
     {2M^{2}\alpha-3Mr(1+\alpha)+r^{2}(1+\alpha)},\\
 E/\mu&=&\frac{M^{2}\alpha-2Mr(1+\alpha)+r^{2}(1+\alpha)}
 {r\sqrt{(1+\alpha)(2M^{2}\alpha-3Mr(1+\alpha)+r^{2}(1+\alpha))}}.
\end{eqnarray}
And the radius of the ISCO is given by
\begin{eqnarray}
 r_{ISCO}/M=2+(1+\alpha)^{-\frac{2}{3}}K^{\frac{1}{3}}
   +(4+\alpha)(1+\alpha)^{-\frac{1}{3}}K^{-\frac{1}{3}},
\end{eqnarray}
where $K=8+\alpha^{2}+(7+\sqrt{5+\alpha})\alpha$. When $\alpha=0$, one will get
\begin{eqnarray}
 r_{ISCO}/M=6,\quad
 L_{ISCO}/M\mu=2\sqrt{3},\quad
 E_{ISCO}/\mu=\frac{2\sqrt{2}}{3},
\end{eqnarray}
for the Schwarzschild black hole. On the other hand, when $a\neq0$, the analytical result can be obtained for the Kerr black hole with $\alpha=0$, see Ref. \cite{Bardeen}. While when $\alpha\neq0$, no analytical result is available. Nevertheless, we can numerically solve Eq. (\ref{ISCOcondition}). The results are listed in Figs. \ref{pRpro} and \ref{pRret} for prograde and retrograde ISCOs, respectively. For the prograde ISCO, both the angular momentum $L_{ISCO}$ and radius $r_{ISCO}$ decreases with the black hole spin $a$ and the parameter $\alpha$. For the retrograde ISCO, we clearly see that the angular momentum $L_{ISCO}$ decreases and the radius $r_{ISCO}$ increases with the spin $a$. However, $L_{ISCO}$ increases and $r_{ISCO}$ decreases with $\alpha$.

\subsection{Equal spin merger}

\begin{figure}
\center{\subfigure[]{\label{afnua}
\includegraphics[width=6cm]{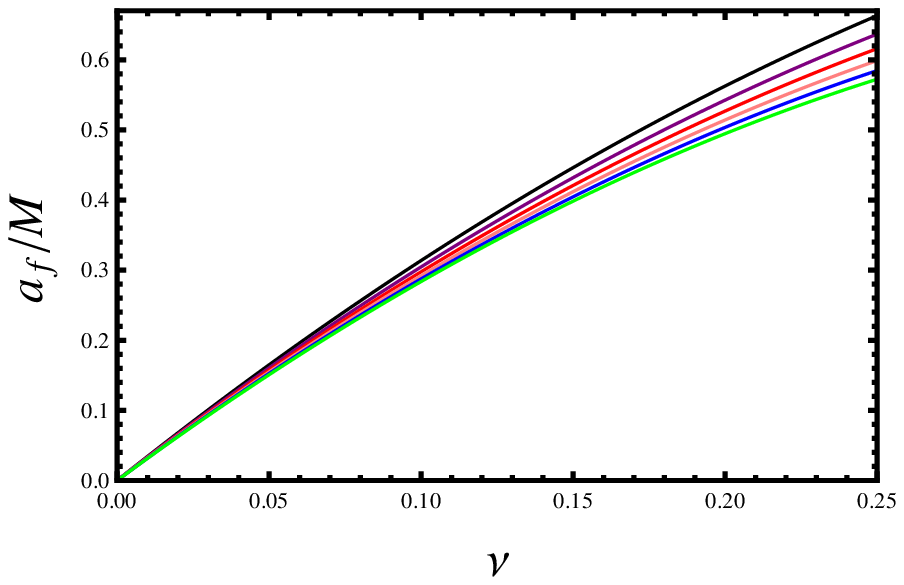}}
\subfigure[]{\label{afnub}
\includegraphics[width=6cm]{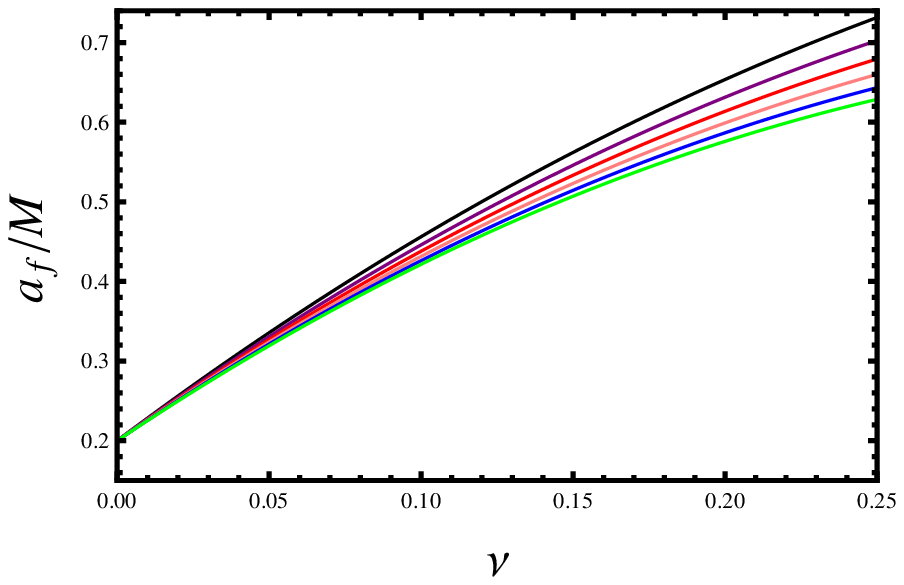}}\\
\subfigure[]{\label{afnuc}
\includegraphics[width=6cm]{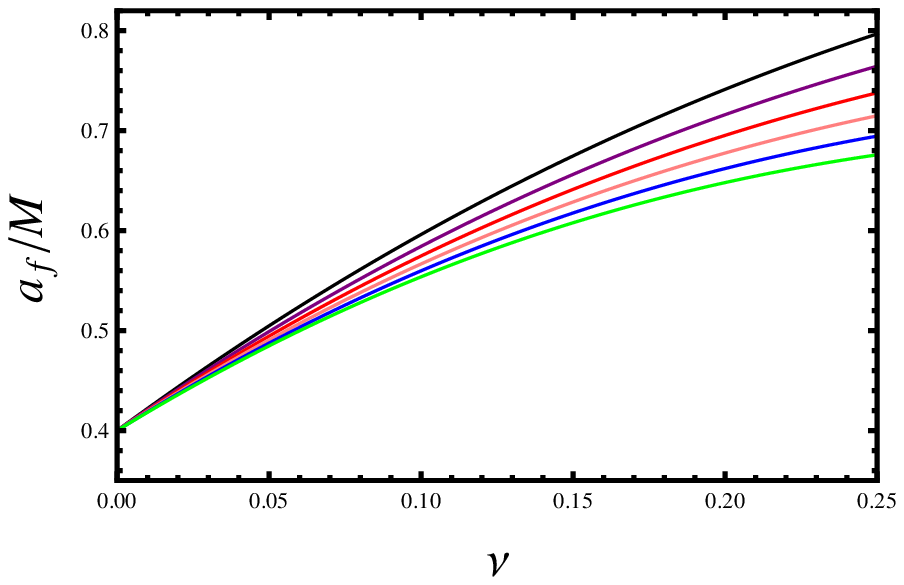}}
\subfigure[]{\label{afnud}
\includegraphics[width=6cm]{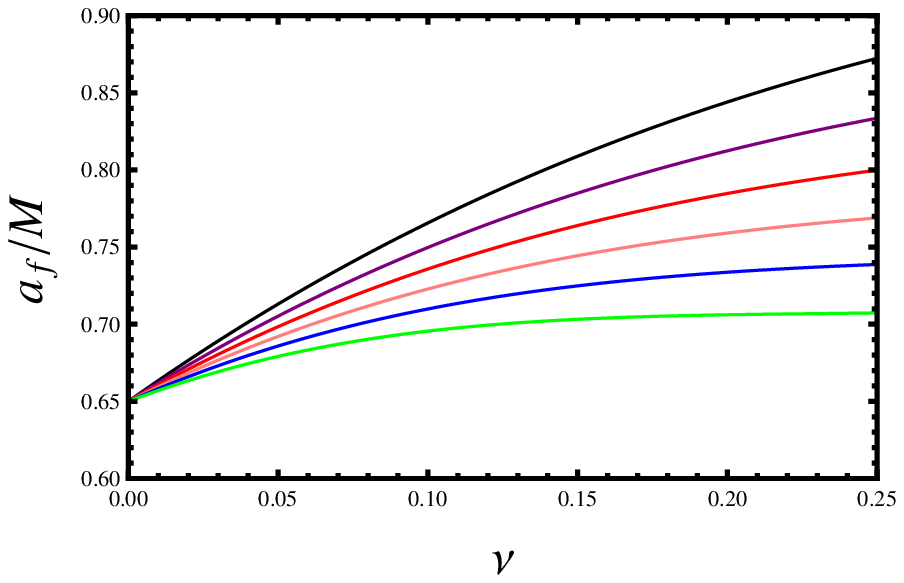}}}
\caption{Final spin $a_{f}$ vs. mass parameter $\nu$ for equal spin merger. (a) $\chi$=0. (b) $\chi$=0.2. (c) $\chi$=0.4. (d) $\chi$=0.65. The parameter $\alpha$=0, 0.2, 0.4, 0.6, 0.8, 1 from top to bottom.}\label{pafnud}
\end{figure}

Based on the study of the ISCO for the black hole, we would like to estimate the final spin parameter of the merger. First, we consider a simple case that the two black holes have the equal spin, i.e., $\chi_{1}=\chi_{2}=\chi$. Therefore, the final spin (\ref{BKL}) will be of the form
\begin{eqnarray}
 a_{f}=L(r_{ISCO}, a_{f})\nu+M(1-2\nu)\chi.
\end{eqnarray}
Given $\nu$ and $\chi$, we can solve the final spin from the above equation. Here we plot the final spin $a_{f}$ as a function of the mass parameter $\nu$ for $\chi$=0, 0.2, 0.4, and 0.65, respectively, in Fig. \ref{pafnud}. For the case of $\chi$=0, it describes a merger of two nonrotating black holes. If these two black hole have the same mass parameter $\nu=0.25$, then we will obtain $a_{f}\approx0.6631$ for the Kerr black hole with $\alpha$=0. Interestingly, this value will decrease when $\alpha$ is nonvanishing. For example, $a_{f}\approx$ 0.6368, 0.6160, 0.5990, 0.5847, 0.5724 for $\alpha$=0.2, 0.4, 0.6, 0.8, 1, respectively. We can also find that the final spin $a_{f}$ increases with $\nu$, while decreases with $\alpha$. If these black holes have initial spins, the result are the similar (see Figs. \ref{afnub}, \ref{afnuc}, and \ref{afnud}).

\begin{figure}
\center{
\includegraphics[width=6cm]{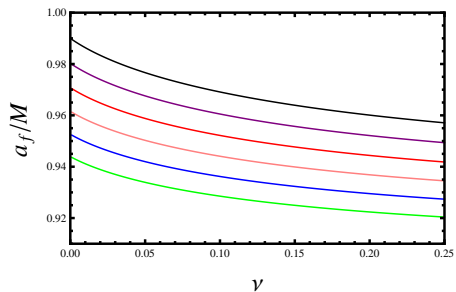}}
\caption{Final spin $a_{f}$ vs. mass parameter $\nu$ for equal spin merger with initial high spin $\chi=0.99\ast\frac{1}{\sqrt{1+\alpha}}$. The parameter $\alpha$=0, 0.02, 0.04, 0.06, 0.08, 0.1 from top to bottom.}\label{afv8}
\end{figure}

On the other hand, we would like to examine the behavior of the final spin $a_{f}$ when the two merged black holes approach its extremal case. For this purpose, we list the final spin $a_{f}$ for $\chi=0.99\ast\frac{1}{\sqrt{1+\alpha}}$ in Fig. \ref{afv8}. From it, one can clearly see that the final spin decreases with the mass parameter $\nu$, which is quite different from that of Fig. \ref{pafnud}. This is a novel result. And it is also consistent with the result given in Refs. \cite{Buonanno,Chatrabhuti}.

\begin{figure}
\center{
\includegraphics[width=6cm]{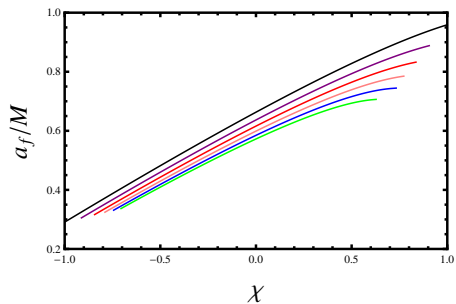}}
\caption{Final spin $a_{f}$ vs. initial spin $\chi$ for equal spin and equal mass black hole merger with $\alpha$=0, 0.2, 0.4, 0.6, 0.8, 1 from top to bottom.}\label{afchi}
\end{figure}

Next, we consider another simple case that the two initial black holes have the same mass, i.e., $\nu=0.25$. Then we present the final spin $a_{f}$ for different initial values of the spin $\chi$ in Fig. \ref{afchi}. Positive or negative $\chi$ denotes that the initial spin is either aligned or antialigned with respect to the initial orbital angular momentum. From the figure, it is clear that the final spin $a_{f}$ increases with $\chi$ from negative to positive values. For fixed $\chi$, the final spin $a_{f}$ decreases with $\alpha$. For example, varying $\alpha$=0, 0.2, 0.4, 0.6, 0.8, 1, $a_{f}(\chi=-0.5)/M\approx$0.4819, 0.4599, 0.4430, 0.4296, 0.4186, 0.4094 and $a_{f}(\chi=0.5)/M\approx$0.8281, 0.7935, 0.7645, 0.7391, 0.7158, 0.6936. Moreover, we can also obtain the result that no matter the initial spin is alignment or antialignment, the final spin of the black hole is always aligned with the initial orbital angular momentum.

\subsection{Unequal spin merger}

\begin{figure}
\center{\subfigure[]{\label{afbetaa}
\includegraphics[width=6cm]{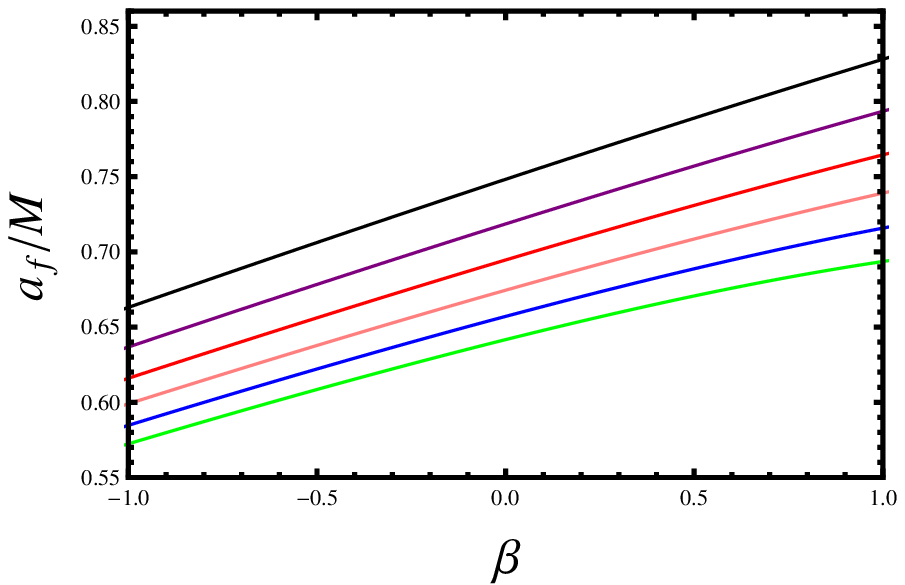}}
\subfigure[]{\label{afbetab}
\includegraphics[width=6cm]{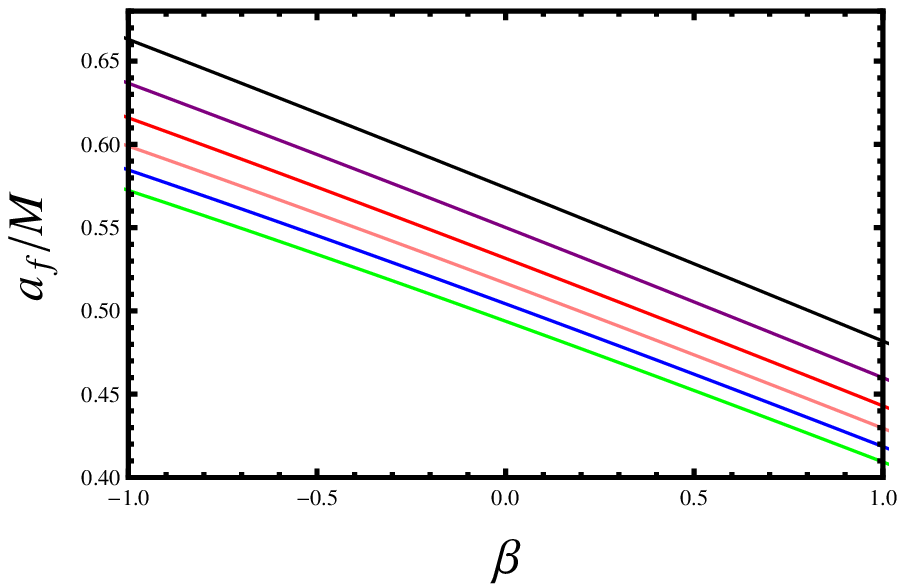}}}
\caption{Final spin $a_{f}$ vs. $\beta$ for unequal spin merger with $\alpha$=0, 0.2, 0.4, 0.6, 0.8, 1 from top to bottom. (a) $\chi$=0.5. (b) $\chi$=-0.5.}\label{pafbetab}
\end{figure}

In this subsection, we would like to consider the unequal spin case but with equal mass. So we set $\chi_{2}=\beta\chi$, $\chi_{1}=\chi$, and $\nu=0.25$. Adopting these values, the final spin will be in the following form
\begin{eqnarray}
 a_{f}=\frac{1}{4}\Big(L(r_{ISCO}, a_{f})+M\chi+\beta M\chi\Big).
\end{eqnarray}
Taking one of the black hole spins as 0.5, and regarding that it is aligned or antialigned with respect to the initial orbital angular momentum, we illustrate the value of the final spin parameter for equal mass black holes in Fig. \ref{pafbetab}. For $\beta$ varies from -1 to 1, we can see that the final spin increases or decreases with $\beta$ for aligned or antialigned case. It is also obvious that the final spin decreases with $\alpha$ for both cases.

\subsection{Generic spin configuration merger}

\begin{figure}
\center{\subfigure[]{\label{achia}
\includegraphics[width=6cm]{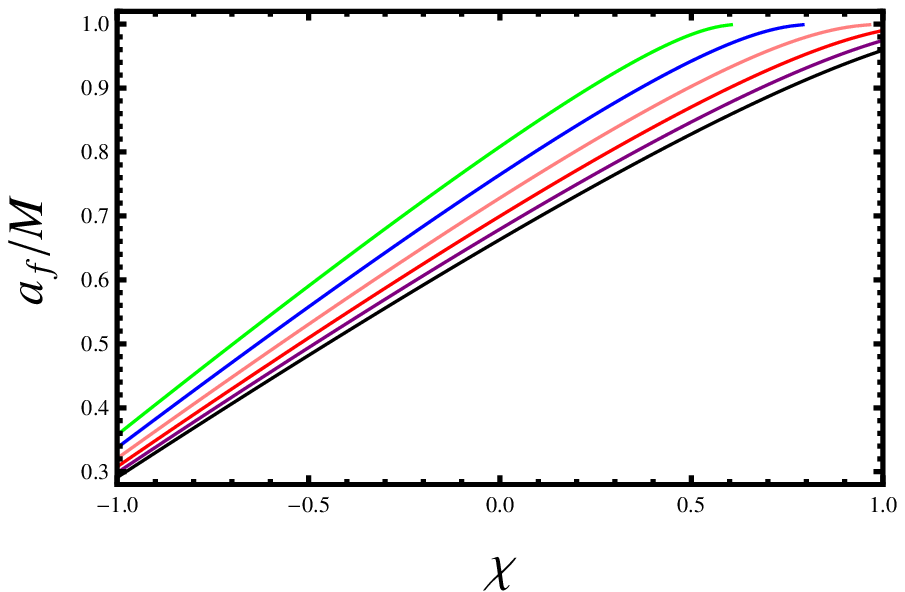}}
\subfigure[]{\label{achib}
\includegraphics[width=6cm]{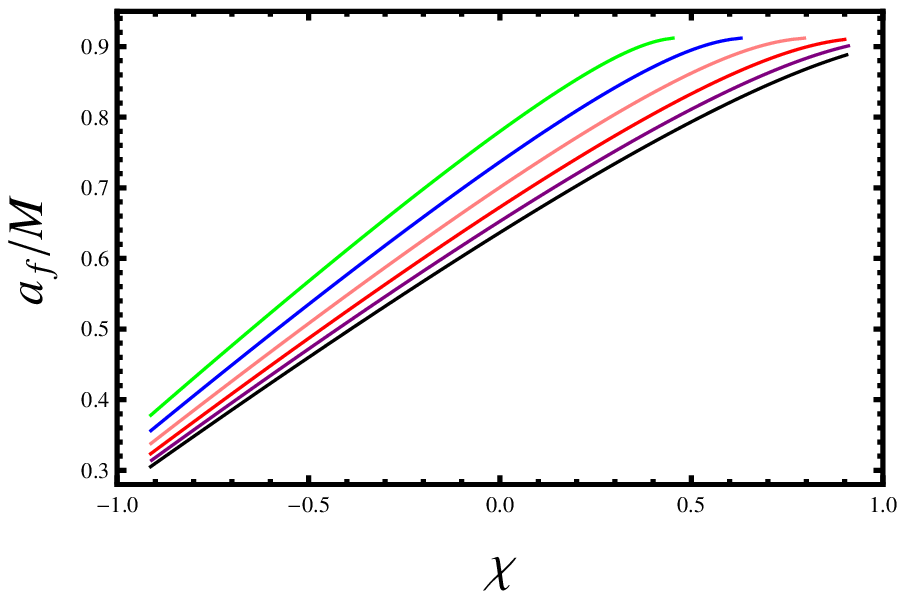}}\\
\subfigure[]{\label{achic}
\includegraphics[width=6cm]{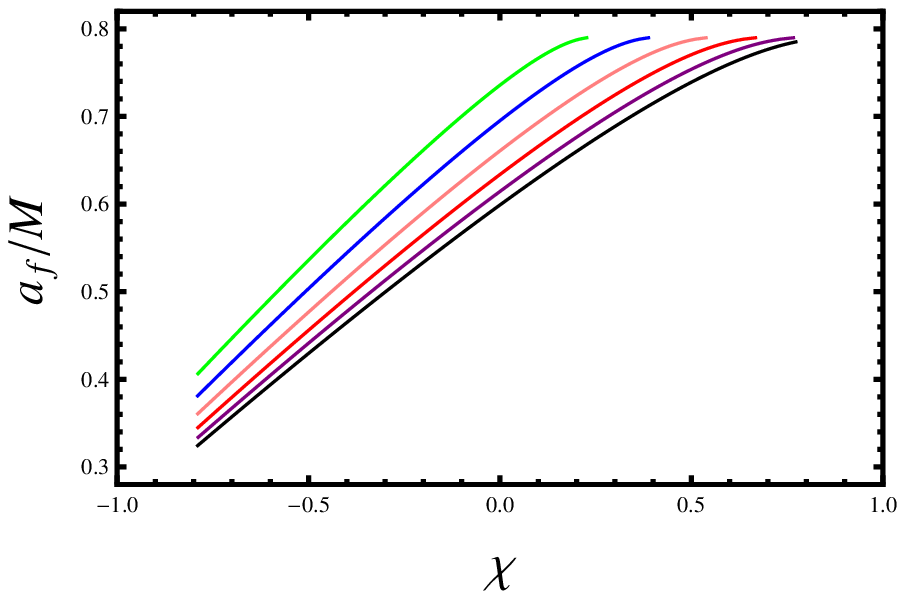}}
\subfigure[]{\label{achid}
\includegraphics[width=6cm]{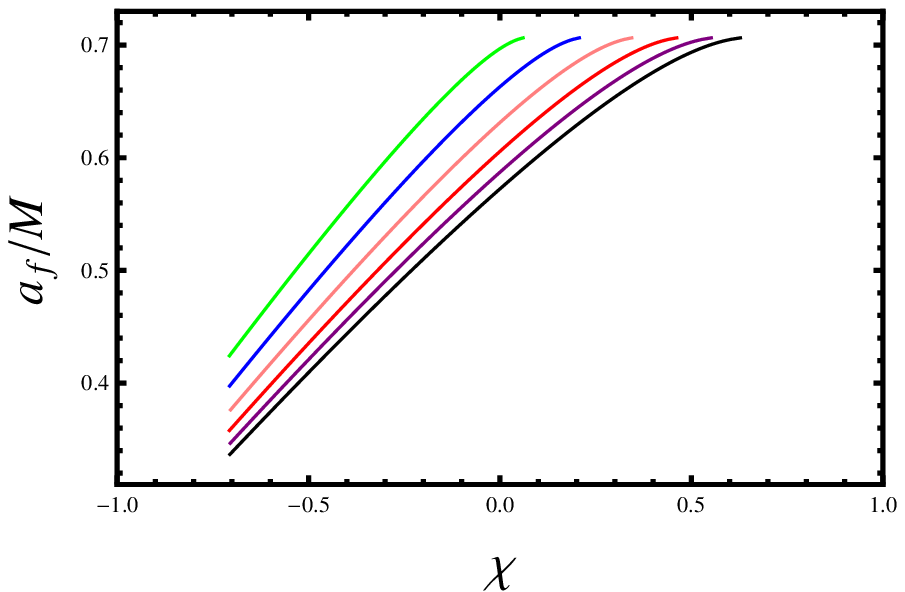}}}
\caption{Final spin $a_{f}$ vs. $\chi$ with $\vartheta$=$0^{\circ}$, $30^{\circ}$, $45^{\circ}$, $60^{\circ}$, $75^{\circ}$, $90^{\circ}$ from bottom to top. (a) $\alpha$=0. (b) $\alpha$=0.2. (c) $\alpha$=0.6. (d) $\alpha$=1. Note that $|\chi|\leq\frac{1}{\sqrt{1+\alpha}}$ for a black hole merger.}\label{pachid}
\end{figure}

Now, we consider a generic spin configuration merger, which means that the orbit plane of the ISCO can be inclined with respect to the final total angular momentum. For this case, the calculation of the orbital contribution to the total angular momentum is required to perform a numerical integration of generic geodesics in the black hole background or use the radial potential for some certain quasiadiabatic spherical orbits \cite{Chandrasekhar}. On the other hand, one can alternatively adopt the fit formula given in Ref. \cite{Hughes}. Here we choose the latter one. Then the orbital angular momentum of the inclined orbit reads \cite{Hughes,Buonanno}
\begin{eqnarray}
 L(\vartheta, a_{f})=\frac{1}{2}(1+\cos\vartheta)L^{pro}(r_{ISCO}^{pro}, a_{f})
  +\frac{1}{2}(1-\cos\vartheta) \Big|L^{ret}(r_{ISCO}^{ret}, a_{f})\Big|,
\end{eqnarray}
where the inclination angle $\vartheta$ measures the angle between the final spin $a_{f}$ and the orbital angular momentum. Let us consider a simple case that these two merged black holes has the same masses and spins. Thus the final spin $a_{f}$ can be solved as
\begin{eqnarray}
 a_{f}=\frac{1}{8}\Big(L^{pro}+|L^{ret}|+4M\chi
   +(L^{pro}-|L^{ret}|)\cos\vartheta\Big).
\end{eqnarray}
Here we show the final spin $a_{f}$ as a function of $\chi$ for different values of $\vartheta$ and $\alpha$ in Fig. \ref{pachid}. From it, we can see that the angle $\vartheta$ has a significant influence on the final spin $a_{f}$. And with the increase of $\vartheta$, the final spin $a_{f}$ increases.

\section{Conclusions}
\label{Conclusion}

In this paper, we studied the main features of the Kerr-MOG black hole merger in a modified gravity. The results show a significant dependence on the dimensionless scalar field parameter $\alpha$.

At first, we explored the property and geodesics for the Kerr-MOG black hole. For the nonrotating case with $\alpha\neq0$, the black hole has two horizons, which is different from the Schwarzschild black hole case. On the other hand, compared to the rotating Kerr black hole, the maximum spin will be reduced by $\alpha$, i.e., $a_{max}/M=\frac{1}{\sqrt{1+\alpha}}$. Then the geodesics was obtained, which also shows an $\alpha$-dependent property.

Based on the null geodesics, we investigated the gravitational waves at the ringdown stage, which can be effectively approximated by the light rings. The real and imaginary parts are, respectively, described by the angular velocity $\Omega_{c}$ and the Lyapunov exponent $\lambda$. Our result implies that, for the prograde orbit, $\Omega_{c}$ increases with $\alpha$, and $\lambda$ increases with $\alpha$ for small $a$ while decreases with $\alpha$ for large $a$. For the retrograde orbit, $\Omega_{c}$ decreases with $\alpha$, while $\lambda$ increases with it.

Next, using the result of the ISCO for the massive particles, we estimated the final black hole spin via the BKL recipe. Several especially interesting cases were explored.

For the case of the equal spin merger, the final spin $a_{f}$ decreases with the dimensionless scalar field parameter $\alpha$. It was also found that the final spin $a_{f}$ increases with $\nu$ for low initial spin. While for the initial spin approaches its maximum value, $a_{f}$ will not increase but decrease with $\nu$, see Fig. \ref{afv8}. So we can expect that there may exist a critical value of $\chi$ that the final spin $a_{f}$ is independent of the mass parameter $\nu$. Certainly, such critical value will be reduced by $\alpha$. Further, if these two merged black hole have equal masses, the final spin will also reduced by $\alpha$. Nevertheless, no matter the initial spins is alignment or antialignment, the final spin of the black hole is always aligned with the initial orbital angular momentum.

For the case of the unequal spin merger, we found that the final spin increases or decreases with $\beta$ for aligned or antialigned case. Moreover, for both the cases, the final spin presents a monotonically decreasing behavior with the increase of the dimensionless scalar field parameter $\alpha$.

We also considered the generic spin configuration merger, which allows the orbit plane has an inclination angle with respect to the final total angular momentum. Then we adopted the fit formula given in Ref. \cite{Hughes} to modify the orbital angular momentum. Considered the two black holes have equal mass and spin, we studied the final spin $a_{f}$ as the initial spin $\chi$ for different inclination angle $\vartheta$ and $\alpha$. The result implies that the final spin $a_{f}$ increases with the inclination angle $\vartheta$.

In summary, our study reveals that for this modified gravity, the black hole merger closely depends on its characteristic parameter, i.e., the dimensionless scalar field parameter $\alpha$. We expected that such influence of $\alpha$ can be determined by the gravitational wave detection in the near future.

\section*{Acknowledgements}
This work was supported by the National Natural Science Foundation of China (Grants No. 11675064, No. 11522541, No. 11375075 and No. 11205074).

\end{document}